# Machine Learning-Enabled Wireless Power Transfer: A Task-Oriented Review of Structural Optimization, State and Parameter Estimation, and Adaptive Control

**Denghao Luo[1], Yiheng Jiang[1], Jiaqi Zhao[2], Yuan Gao[3,*], Changsong Cai[1,*], Weiyi Huang[2], and Yun Zhang[4]**

[1] School of Electrical Engineering and Automation, Wuhan University, Wuhan, China
[2] Faculty of Electrical and Control Engineering, Liaoning Technical University, Liaoning, China
[3] School of Engineering, University of Leicester, Leicester, UK
[4] School of Electrical and Information Engineering, Tianjin University, Tianjin, China

* Correspondence author(s); E-mail(s): yuan.gao@leicester.ac.uk; changsongcai@whu.edu.cn.

**Highlights:**

- A task-oriented review maps machine learning methods to core WPT tasks.
- Supervised, reinforcement, hybrid, and physics-informed methods are compared.
- The review spans structural design, parameter and state estimation, and control.
- Data, generalization, safety, and deployment gaps guide future WPT research.

**Abstract:** Wireless power transfer (WPT) systems must maintain efficient and safe energy delivery under variations in geometry, coupling, load, alignment, and operating conditions. Machine-learning methods are increasingly applied to accelerate structural design, estimate operating parameters and states, and support adaptive control and scheduling. This review develops a task-oriented framework organized around three interconnected domains: structural design and optimization, parameter and state estimation, and control and optimization. Supervised learning, reinforcement learning, and hybrid or physics-informed methods are compared in terms of data requirements, dependence on physical models, information sources, computational requirements, accuracy, online response, application scope, and optimization objectives. Supervised learning provides rapid surrogate modeling, regression, and classification when representative labeled data are available. Reinforcement learning addresses sequential control and scheduling but introduces challenges in reward design, safe exploration, and simulation-to-reality transfer. Hybrid methods integrate physical knowledge with data-driven models to improve data efficiency and physical consistency. The principal barriers to engineering deployment include limited and biased datasets, weak out-of-distribution generalization, insufficient physical consistency, unsafe online adaptation, embedded-resource constraints, and the absence of standardized benchmarks. Future research should prioritize uncertainty-aware and physics-informed learning, safe simulation-to-reality transfer, lightweight edge deployment, distributed coordination, reproducible evaluation, and digital-twin-assisted development.

## 1. Introduction

### 1.1 Wireless Power Transfer Technology: Background and Application Requirements

The accelerating electrification of transport and industry, together with the growing deployment of intelligent and autonomous systems, is increasing demand for power-delivery solutions that are flexible, safe, and suitable for continuous operation in dynamic environments. Although wired power-delivery systems remain highly efficient and technologically mature, their reliance on physical connectors can introduce maintenance and automation constraints in applications involving frequent connection and disconnection, motion, or sealed interfaces. Wireless power transfer (WPT) provides a contactless alternative and has therefore attracted increasing attention for next-generation energy-delivery systems [1].

WPT has been investigated in electric transportation, unmanned aerial vehicles (UAVs), industrial automation, biomedical implants, underwater equipment, and consumer electronics. In electric-vehicle applications, it can support automated stationary charging and dynamic charging concepts, while coupler design remains central to transfer efficiency and tolerance to misalignment [2,3]. For UAVs, WPT offers a potential route for energy replenishment, whereas in-flight charging remains a specialized research direction [4]. In industrial automation, WPT can enable automated charging of automated guided vehicles (AGVs) [1]. It can also supply implantable biomedical devices, including pacemakers [5], and enable contactless recharging of underwater equipment and autonomous underwater vehicles [6,7]. In consumer electronics, WPT can reduce reliance on physical charging connectors and support compact product designs [1].

WPT research is increasingly extending beyond static applications operated under fixed conditions toward dynamic charging, larger-scale deployment, and more demanding operating environments [8]. This evolution introduces a set of closely coupled technical challenges.

1) Structural design and optimization: Magnetic couplers for dynamic WPT must balance transfer efficiency, misalignment tolerance, electromagnetic compatibility, and size and weight constraints. In dynamic electric vehicles (EVs) charging, vehicle motion changes coil alignment and, consequently, the coupling condition. Conventional design workflows based on finite-element analysis(FEA) and iterative parameter sweeps can therefore make rapid engineering iteration computationally demanding [3].

2) State and parameter estimation: Variables such as coil position, mutual inductance, and load are not always directly measurable and must often be inferred from electrical measurements. The accuracy of purely model-based estimators can be affected by measurement noise, model mismatch, misalignment, and time-varying operating conditions. Recent studies have explored machine-learning-based approaches for position estimation, parameter prediction, and foreign-object detection in WPT systems [9].

3) Control and operational optimization: Dynamic WPT operation requires timely adjustment of operating frequency, impedance matching, and power transfer in response to variations in coupling and load. Fixed-gain or conventionally tuned PI controllers may require retuning and can exhibit reduced robustness when coupling conditions vary substantially [10].

4) System-level integration: WPT performance depends on the coordinated design of magnetic couplers, compensation networks, power converters, sensing functions, and control strategies. Their interdependent objectives and constraints make isolated component-level optimization insufficient for achieving optimal system-level performance. Recent machine-learning(ML)-enabled studies illustrate the value of linking offline parameter optimization with online adaptive control, while full system-level co-design remains an open research direction [11].

These coupled challenges motivate the introduction of ML-based methods into WPT research. Supervised learning, reinforcement learning, and hybrid or physics-informed methods can complement analytical and model-based approaches by accelerating design evaluation, supporting state and parameter inference, and enabling adaptive operation under changing conditions. Accordingly, AI has become an increasingly important research direction for next-generation WPT systems.

### 1.2 Limitations of traditional design and control methods

Before machine-learning methods were introduced into WPT design and operation, analytical models, FEA, and model-based control constituted the principal engineering tools. Magnetic-coupler design commonly combines FEA with parameter sweeps or population-based optimization. Parameter sweeping may require thousands of FEA evaluations [104]. Although genetic algorithms and particle swarm optimization (PSO) reduce reliance on exhaustive enumeration [102,107], candidate evaluation remains computationally expensive, and algorithm settings and penalty coefficients are often problem dependent [108,112].

Analytical electromagnetic models offer greater computational efficiency when their underlying assumptions are satisfied. For example, magnetic-vector-potential-based formulations can calculate mutual inductance accurately for prescribed coil geometries [117]. At the circuit level, compensation networks such as S-S, LCC-S, and LCC-LCC are designed around specific resonance and impedance-matching conditions [121]. Component tolerances, aging, and temperature-induced capacitance drift can cause detuning [122,127], while variations in coupling and load resistance alter the reflected impedance and transfer efficiency [124]. Fixed-parameter networks may therefore require tolerance-oriented design or active tuning.

At the control level, conventional PID controllers can exhibit degraded regulation under variations in load, circuit parameters, and magnetic coupling [132]. Sliding-mode and model predictive controllers improve dynamic performance but rely on explicit system models whose accuracy affects control performance [140,141]. Conventional WPT workflows therefore involve a trade-off among model fidelity, computational cost, and adaptability.

### 1.3 Machine Learning-Enabled Wireless Power Transfer: Motivation and Current State

As WPT systems move toward higher power density, wider spatial coverage, and dynamic operation, system design, state estimation, and control must satisfy increasingly demanding performance requirements. Conventional model-based workflows can incur substantial computational cost and may require model revision or controller retuning as operating conditions change.

Recent studies demonstrate increasing methodological and application diversity. The learning methods applied to WPT now include feedforward backpropagation neural network (BPNN) for real-time impedance matching [14], deep neural network (DNN) for predicting efficiency enhancement in automotive couplers [15], and reinforcement learning for voltage regulation, joint charging and routing, and real-time energy scheduling [16,23,48]. Hybrid approaches further combine neural networks with evolutionary or swarm-based optimization, while physics-informed models incorporate electromagnetic constraints into the learning process [17,153,154]. Applications have also expanded beyond EV charging to underwater WPT and wearable-device charging [19-21], and optimization formulations increasingly consider multiple competing performance objectives [19]. In parallel, research is progressing from offline design toward real-time prediction, adaptive control, and intelligent scheduling [14,16,48].

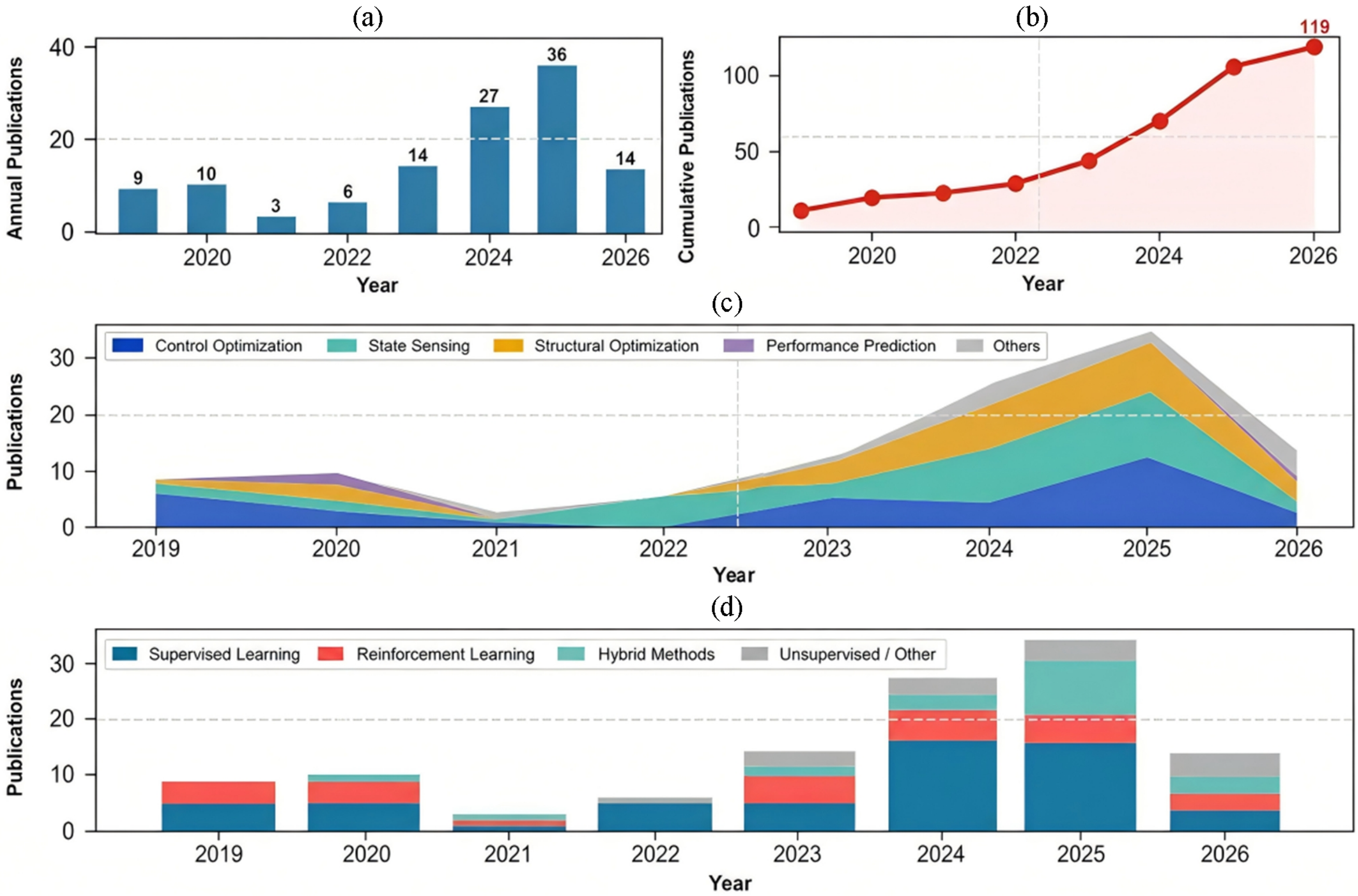

**Figure 1.** Publication trends in recent years. (a) annual publications; (b) cumulative publications; (c) evolution of research tasks; and (d) annual distribution of learning categories..

Figure 1 summarizes the publication trends based on literature retrieved using "machine learning" and "wireless power transfer" as search terms. The literature collected for this review reveals a marked increase in research on AI-enabled WPT after 2022. As shown in Figure 1(a), annual publication output rose from 14 papers in 2023 to 27 in 2024 and 36 in 2025, with the 2025 total reaching 3.6 times the 2020 level. As shown in Figure 1(b), the corresponding cumulative total reached 105 publications by 2025, indicating that the integration of learning-based methods into WPT has developed from a relatively limited research topic into a rapidly expanding area. The growth in publication volume has been accompanied by broader methodological and task coverage. As shown in Figure 1(c), the literature now spans structural optimization, state estimation, performance prediction, and control optimization, with particularly strong growth in state estimation and control-oriented research after 2022. As shown in Figure 1(d), supervised learning remains the most widely adopted category, while reinforcement learning and hybrid methods have become increasingly prominent in recent studies. These trends suggest that AI-enabled WPT research is extending beyond isolated prediction or design problems toward multiple interconnected stages of system development and operation. This increasing diversity provides the basis for the task-oriented classification and comparative analysis adopted in this review.

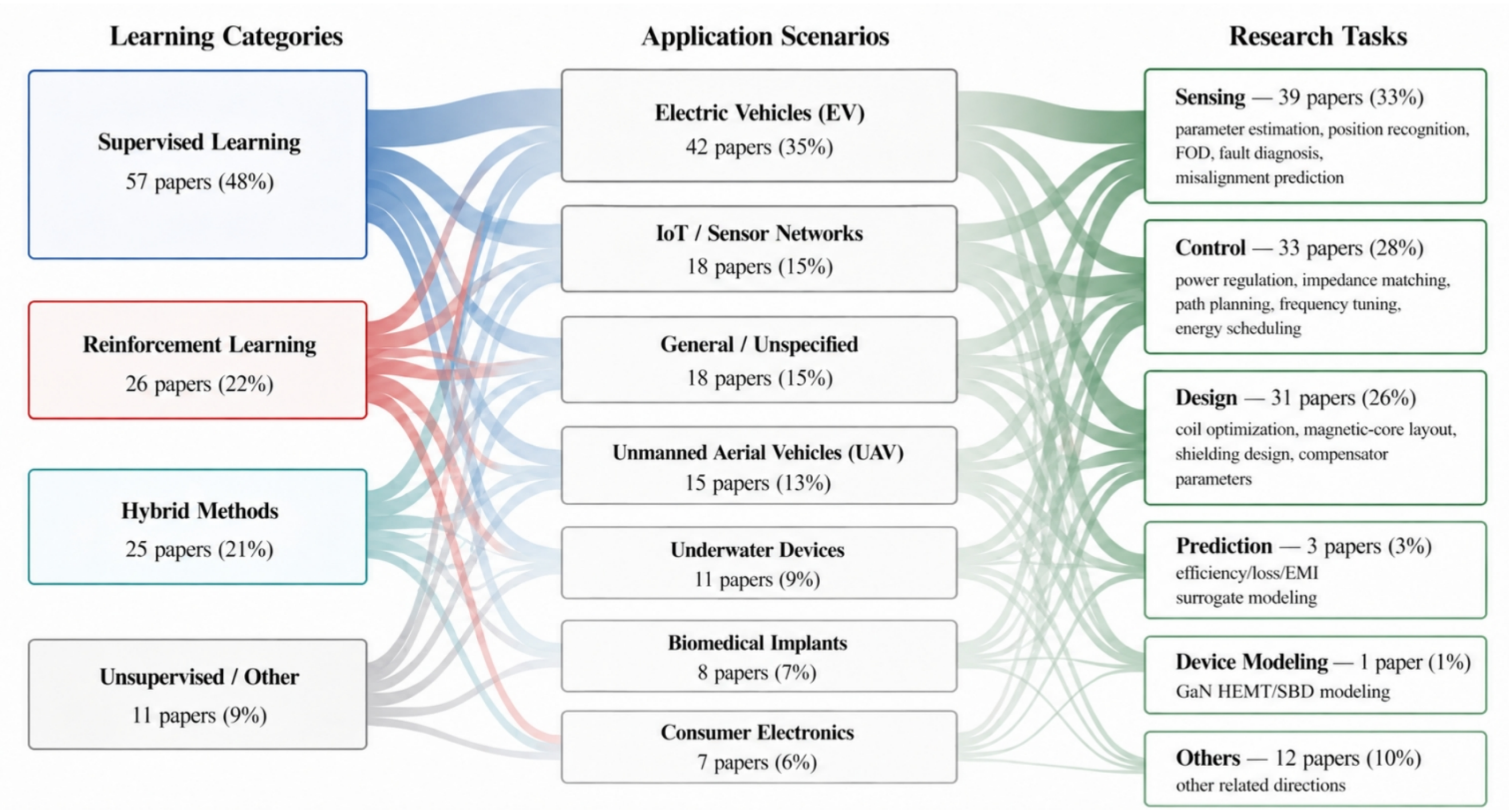

**Figure 2.** Sankey diagram of learning categories, application scenarios, and research tasks in AI-enabled WPT

Figure 2 presents the methodological and thematic composition of the reviewed literature. Supervised learning constitutes the largest methodological category, comprising 57 papers (48%), followed by reinforcement learning with 26 papers (22%), hybrid methods with 25 papers (21%), and unsupervised learning and other approaches with 11 papers (9%). Electric vehicles represent the largest application domain, accounting for 42 papers (35%). Internet of Things (IoT) and sensor networks and general or unspecified applications each account for 18 papers (15%), while the remaining studies focus on UAVs, underwater devices, biomedical implants, and consumer electronics. In terms of research tasks, sensing and state estimation form the largest category, comprising 39 papers (33%), followed by control and optimization with 33 papers (28%) and structural design and optimization with 31 papers (26%). Overall, supervised learning dominates the methodological landscape, while the literature is concentrated in three principal task areas: sensing and state estimation, control optimization, and structural design optimization.

The connections shown in Figure 2 further suggest recurring associations between learning categories and WPT research tasks. In structural design and optimization, supervised surrogate models and physics-informed or meta-model-based frameworks can reduce reliance on repeated full-order electromagnetic simulations and accelerate the modeling and optimization of magnetic couplers [106,153,154]. In sensing and state estimation, data-driven models establish nonlinear mappings between measurable electrical responses and position-, coupling-, or operating-condition-related variables [9,21]. Reinforcement learning supports adaptive regulation, dynamic tuning, charging coordination, and energy scheduling by learning sequential decision-making policies through interactions with the operating environment [16,23,48]. Hybrid methods integrate data-driven models with optimization algorithms or physical knowledge to improve computational efficiency, data efficiency, and physical consistency [17,153,154]. These methodological–task associations provide the basis for the task-oriented framework adopted in this review, through which the applicability, limitations, and engineering readiness of different learning methodologies are systematically compared.

Existing reviews have examined WPT from several complementary perspectives. Alabsi et al. [1] surveyed WPT technologies, application domains, and wireless-charging scheduling schemes for wireless sensor networks, whereas Liu et al. [8] reviewed near- and far-field power-transfer mechanisms, compensation and performance-enhancement techniques, and representative applications. These broad reviews provide extensive coverage of the WPT technology landscape but devote limited attention to how learning methods address specific engineering tasks. Focusing on EV wireless charging, Yadav and Bera [3] critically reviewed electromagnetic coil and magnetic coupler design, emphasizing coil geometry, materials, shielding, power-transfer efficiency, misalignment tolerance, and practical engineering trade-offs. Ali et al. [12] reviewed artificial-neural-network(ANN)-based WPT optimization, covering ANN models and their applications in performance prediction, coil design, misalignment compensation, energy management, and power-converter and fault analysis. Amadei et al. [9] reviewed a broader range of ML applications in WPT, including coupler design, parameter and efficiency prediction, frequency tuning, coil position estimation, misalignment compensation, and foreign object detection (FOD). Haddad et al. [13] reviewed ML-based optimization for simultaneous wireless information and power transfer (SWIPT), with particular attention to beamforming, relay selection, power and receiver control, and resource allocation.

These studies provide valuable technology-, component-, application-, and algorithm-oriented perspectives. However, among the reviews considered here, to the best of our knowledge, none of the reviewed studies a common task-oriented framework that compares supervised learning, reinforcement learning, and hybrid or physics-informed methods across structural design and optimization, online parameter and state estimation, and adaptive control and scheduling. Comparative assessments across these stages also remain limited with respect to data requirements, dependence on physical models, robustness and generalization under changing operating conditions, safety constraints, real-time capability, and practical deployment readiness. Consequently, selecting an appropriate learning methodology for a specific WPT task remains difficult, as does determining whether performance improvements reported in offline studies can be transferred to practical WPT systems.

The literature reviewed in this paper is organized into three task categories: structural design and optimization, state estimation, and control and scheduling. Within each domain, supervised learning, reinforcement learning, and hybrid or physics-informed methods are compared in terms of data requirements, physical-model dependence, computational efficiency, real-time performance, robustness, and practical deployment readiness. This task-oriented framework clarifies the applicability and limitations of different learning methodologies, identifies shared bottlenecks and underexplored links across the WPT system lifecycle, and provides a structured basis for methodology selection and future research.

### 1.4 Scope and Organizational Structure

This review first examines the conventional methods used in WPT system design, state estimation, and control, with particular attention to their limitations in computational efficiency, model dependence, and adaptability to changing operating conditions. It then introduces and categorizes the principal learning algorithms currently applied to WPT, including supervised learning, reinforcement learning, and hybrid or physics-informed methods. Building on this methodological foundation, recent studies are systematically reviewed across three core WPT tasks: structural design optimization, state estimation, and control optimization. The applicability, strengths, and limitations of different learning methods are compared in terms of data requirements, computational complexity, physical-model dependence, real-time capability, robustness, and practical deployment readiness. Finally, the review discusses the principal challenges facing intelligent WPT systems, including limited access to representative data, inadequate generalizability, insufficient physical consistency, and resource constraints associated with edge deployment. Future research directions are also outlined, including physics-informed deep learning, lightweight model development, multi-agent coordination and control, and digital-twin-enabled WPT. Through this progression from conventional approaches to learning methodologies, task-specific applications, and future research opportunities, the review provides a structured account of the development and engineering potential of intelligent WPT systems.

## 2. Traditional design and control methods for wireless power transfer

Over the past few decades, WPT has matured into a well-established engineering field grounded in circuit theory and electromagnetic field theory. Prior to the recent integration of data-driven approaches, WPT system design and operation relied primarily on analytical modeling, FEA, model-based estimation, heuristic optimization, and classical and model-based control. Figure 3 summarizes the roles and interconnections of these conventional approaches. This section first reviews traditional system analysis and design methods, including parameter sweeping, genetic algorithms, PSO, model-based parameter and state estimation, compensation-network design, and parity-time (PT)-symmetric circuits. It then examines conventional linear and nonlinear control methods, focusing on their performance under nonlinear, time-varying, and uncertain operating conditions. Finally, it summarizes the common limitations of traditional WPT methods in terms of computational cost, dependence on accurate physical models, and adaptability to changing operating conditions.

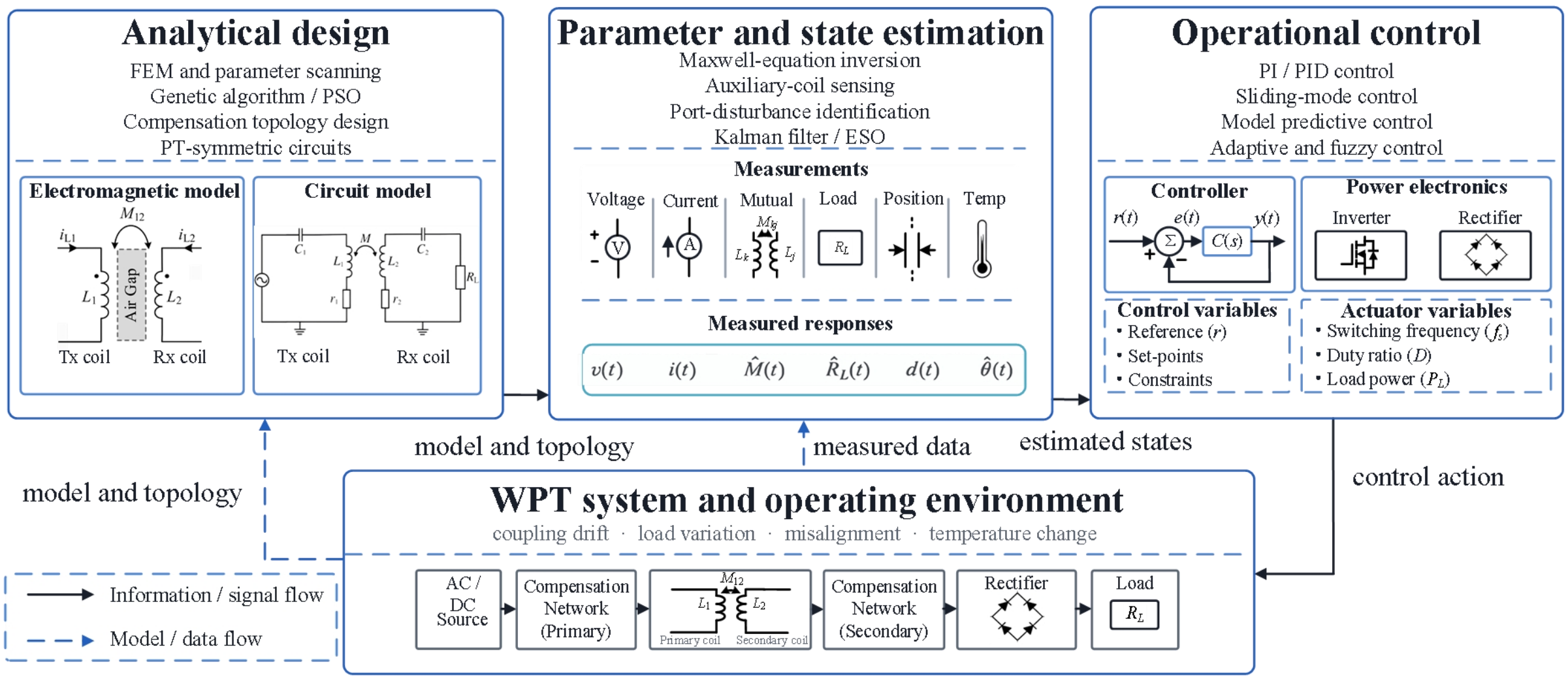

**Figure 3.** Schematic Overview of Traditional WPT Design and Control Workflows.

## 2.1 Traditional Analytical Design Methods

Conventional magnetic-coupler design relies on analytical modeling, FEA, parameter sweeps, and population-based optimization. Analytical subproblem methods and design-of-experiments-assisted searches can reduce the number of full three-dimensional FEA evaluations [101,103], while parameterized meshing reduces the remeshing overhead associated with FEA-based parameter sweeps [104]. Nevertheless, optimization involving multiple geometric and material variables may still require repeated electromagnetic simulations and incur substantial computational cost [103,104]. Nondominated sorting genetic algorithm II (NSGA-II) and PSO replace exhaustive enumeration with guided population-based searches and have been applied to multi-objective magnetic-coupler design and dynamic EV charging [102,107]. Their convergence, however, remains dependent on population diversity and algorithm settings [110], while penalty-function-based constraint handling can be sensitive to problem-dependent penalty coefficients [111,112]. Analytical expressions have also been developed and experimentally validated to quantify transfer efficiency under lateral and angular coil misalignment [150].

Compensation networks such as S–S, LCC–S, and LCC–LCC are designed to achieve specific resonance, impedance-matching, and output characteristics [121]. In practical systems, capacitor tolerances, aging, and temperature variations can cause detuning [122,127], while changes in coupling and load conditions alter the reflected impedance and power-transfer efficiency [124]. Active tuning has therefore been implemented using PWM-controlled equivalent inductance, switched-capacitor-based reactance adjustment, PSO-based tuning of reconfigurable matching networks, and impedance-decoupling schemes for dual-side capacitance drift [123,125-127]. These methods can restore resonance, regulate output power, or maintain high transfer efficiency under changing operating conditions. However, they generally require additional sensing, control, switching, or power-electronic resources.

PT-symmetric WPT provides an alternative means of maintaining efficient power transfer over a prescribed coupling range by balancing effective gain and loss [128-130]. Recent studies have extended this framework through fixed-frequency dual-side active tuning, exceptional-point-based efficiency maximization, high-order compensation networks, and multi-degree-of-freedom misalignment-tolerant structures [128-131]. Nevertheless, system performance remains dependent on the achievable PT-symmetric operating region and the practical realization of the required active gain and tuning mechanisms.

These conventional WPT design approaches combine physics-based modeling with numerical and population-based optimization. Although they provide physical interpretability and follow well-established engineering procedures, their computational efficiency and adaptability are limited by repeated electromagnetic simulations and problem-dependent parameter tuning. These limitations have motivated the development of surrogate and learning-based design methods, which are reviewed in subsequent sections.

## 2.2 Traditional Control Methods

Closed-loop control in WPT systems requires both reliable estimates of operating conditions and an appropriate control law. Because quantities such as mutual inductance, load conditions, receiver position, and internal system states cannot always be measured directly, parameter and state estimation is closely integrated with feedback regulation. This section therefore reviews conventional control architectures from two related perspectives: the acquisition of control-relevant state information and the subsequent regulation of system outputs.

Conventional parameter and state estimation relies primarily on analytical models, sensing coils, calibration maps, and recursive estimators. Magnetic-vector-potential formulations can accurately calculate mutual inductance for prescribed coil geometries [117], while coupled electromagnetic and circuit models can characterize the effects of different transmission media [118]. Receiver position can be inferred from equivalent-mutual-inductance patterns obtained using sensing intermediate coils [114] or estimated by combining receiver-current measurements with vehicle-dynamics information through an unscented Kalman filter [119]. Kalman filtering has also been integrated with model predictive control to provide real-time state estimation and output regulation in multioutput WPT systems [120]. These approaches provide physically interpretable estimates, but their accuracy depends on the fidelity of the underlying model or calibration map and on the quality of the measured signals. Temperature-induced increases in winding resistance can alter transfer efficiency and the admissible coupling range [116], illustrating the importance of accounting for operating-condition-dependent parameters.

Using the estimated parameters and states as feedback information, conventional controllers regulate output voltage, current, or transferred power under variations in load, circuit parameters, and magnetic coupling. PI and PID controllers remain widely used because of their simple structures and low implementation requirements. Their fixed gains, however, may result in degraded regulation when the system operates far from its nominal condition or is subject to substantial parameter variations [132,136,138]. Robust $H_\infty$ linear quadratic regulator, and full-state-feedback controllers have been developed to improve disturbance rejection and transient performance within prescribed operating ranges [132,133,135,137]. These methods require an appropriate dynamic model and, in the case of full-state feedback, measurable or estimated state variables. Observer-based implementations can reduce the number of directly measured signals while retaining the information required by the controller [134].

Nonlinear and model-based control strategies have also been investigated to address the resonant,

coupled, and operating-point-dependent dynamics of WPT systems [140-142]. Sliding-mode control (SMC) provides rapid dynamic response and robustness to bounded disturbances, and discrete SMC has been implemented through frequency modulation in inductive power transfer systems [140]. Conventional SMC may nevertheless exhibit chattering, motivating composite passivity-based and terminal sliding-mode formulations [147]. Other control strategies address specific operating characteristics. Linear-parameter-varying control accommodates rectifier conduction-mode and load variations [143], fuzzy-logic control supports transitions between constant-current and constant-voltage charging modes [144], and passivity-based control has been applied to output regulation under time-varying coupling conditions [145]. Switching nonidealities, including inverter dead time, may require additional dedicated compensation mechanisms [148].

Model predictive control (MPC) determines the control action by predicting system behavior over a finite horizon and optimizing a specified objective. In WPT systems, MPC has been used to enforce input constraints, regulate output variables, and coordinate tracking accuracy, transfer efficiency, and soft-switching requirements [141,146,149]. Its practical performance depends on the adequacy of the prediction model and the computational resources available for online optimization. Primary-side parameter estimation can improve adaptation to changing mutual inductance and reflected-load conditions [151], while low-order data-driven models can reduce the digital implementation burden associated with high-order circuit models and communication delays [152]. Finite-control-set MPC has also been investigated as a model-based control strategy for WPT systems [139].

The available evidence indicates that these conventional methods offer complementary capabilities rather than a universal control solution. PI and PID controllers provide low implementation complexity but require careful tuning across changing operating conditions. Robust and state-feedback controllers depend on suitable dynamic models and state information, SMC may require additional measures to suppress chattering, and MPC must balance model fidelity, constraint handling, and online computational cost. The following subsection summarizes the common limitations of these conventional methods in terms of model dependence, computational cost, and adaptability.

## 2.3 Common Limitations in Traditional Methods

Across conventional WPT design, estimation, and control methods, recurring trade-offs arise among model fidelity, computational cost, and adaptability. Physics-based analytical models provide interpretability but become increasingly difficult to formulate when electromagnetic, circuit, thermal, and operating effects are strongly coupled. FEA provides high-fidelity electromagnetic evaluation, but repeated simulations can substantially increase the cost of design-space exploration and iterative optimization. Reduced-order and lumped-parameter models improve computational efficiency, but may omit geometric details, edge effects, material nonlinearities, and high-frequency losses [153].

Conventional controllers can provide effective regulation within their modeled operating regions, but their performance may deteriorate when parameter variations, disturbances, or unmodeled dynamics exceed the assumptions adopted during controller design. Moreover, most conventional controllers do not explicitly learn reusable nonlinear mappings or control policies from accumulated operating data; instead, uncertainty is generally addressed through robust or adaptive design and online state or parameter estimation.

Machine learning provides complementary tools for addressing these limitations. Surrogate and meta-models can approximate computationally expensive electromagnetic evaluations, thereby accelerating magnetic-coupler design and multi-objective optimization [106,154]. Physics-informed neural networks further incorporate governing equations and electromagnetic constraints into the learning process, improving physical consistency while reducing reliance on repeated FEA simulations [153]. In state estimation and operational control, learning-based methods can exploit measured data to characterize nonlinear and time-varying relationships that are difficult to represent using fixed analytical models. These methods should nevertheless be regarded as complements rather than universal replacements for physics-based approaches, because data requirements, generalization, physical consistency, safety, and deployment cost remain important evaluation criteria.

Table 1 establishes the task-oriented framework adopted in this review. It summarizes the representative variables, learning roles, and evaluation questions associated with structural design optimization, parameter and state estimation, and operational control. The subsequent sections use this framework to compare the applicability, limitations, and engineering readiness of different learning methodologies.

**Table 1. Application layers, learning roles, and key review questions for ML-enabled WPT**

| Task Topics | Typical variables | Learning role | Key review question |
| --- | --- | --- | --- |
| Structural design optimization | Coil geometry, ferrite layout, shielding, compensation network, frequency | Surrogate modeling; optimizer acceleration; multi-physics approximation | Does the learned model preserve electromagnetic, thermal, and manufacturing constraints? |
| Parameter and state estimation | Voltage, current, phase, impedance, harmonics, S-parameters, auxiliary-coil signals | Regression, classification, inverse mapping, uncertainty estimation | Does the estimator generalize across offset, load, topology, temperature, and medium? |
| Control and scheduling | Duty ratio, phase shift, frequency, coil activation, beam direction, power allocation | Policy learning, adaptive tuning, safe mode selection, long-horizon scheduling | Can the controller improve performance without violating hardware safety limits? |

## 3. Machine Learning Algorithm Classification and Comparative Analysis

The preceding analysis identified three recurring limitations of conventional WPT design, estimation, and control methods: high computational cost, dependence on predefined models, and limited adaptability to changing operating conditions. These limitations become particularly important in high-dimensional design spaces, time-varying operating environments, and multi-objective optimization problems involving coupled constraints. They reflect fundamental trade-offs rather than universal failures. High-fidelity analytical and finite-element models can be computationally expensive to construct and evaluate, whereas reduced-order or linearized models may lose accuracy when the system operates outside the conditions assumed during model development.

Machine-learning methods provide complementary tools by learning nonlinear relationships from simulation or experimental data and by supporting adaptive decision-making. However, supervised learning, reinforcement learning, and hybrid or physics-informed methods differ substantially in their data requirements, dependence on physical models, training and inference procedures, deployment characteristics, and engineering limitations. These differences complicate direct comparison across studies and make it difficult to determine which learning category is appropriate for a particular WPT task.

Section 3 therefore first classifies the literature into supervised learning, reinforcement learning, and hybrid or physics-informed methods, and then compares these categories in terms of data and model dependence, information sources, computational requirements, accuracy, response speed, application scope, and optimization objectives. This analytical structure supports the task-oriented review in Section 4, which examines the application of machine learning to structural optimization, parameter and state estimation, and control and optimization. Table 2 summarizes the representative WPT tasks, principal strengths, and major limitations of the three learning categories emphasized in this review.

**Table 2.** Representative WPT tasks, strengths, and limitations of major learning categories.

| Category | Best-suited WPT tasks | Advantages | Limitations |
|---|---|---|---|
| Supervised learning | Surrogate design, state estimation, FOD/fault recognition, device modeling | Fast inference; mature training workflow; suitable for embedded estimation | Requires labeled data; weak extrapolation; limited physical interpretability |
| Reinforcement learning | Dynamic control, frequency tuning, transmitter selection, UAV/IoT scheduling | Handles sequential decisions and delayed rewards; can optimize long-term objectives | Training cost, reward design, safe exploration, and sim-to-real transfer are difficult |
| Hybrid and physics-informed methods | Physics-informed identification, surrogate-assisted optimization, adaptive PI/MPC support | Combines interpretability with data-driven speed; often more deployable | Requires careful model coupling and validation across physical regimes |

## 3.1 Overall Algorithm Classification

Based on the literature reviewed in this study, machine-learning methods applied to WPT are organized into three principal categories: supervised learning, reinforcement learning, and hybrid intelligent methods. These categories are reviewed in terms of their underlying principles, representative algorithms, typical WPT applications, and engineering limitations.

### 1) Supervised learning

Supervised learning constitutes the largest methodological category in the reviewed corpus, accounting for approximately 48% of the included studies. It uses labeled input-output data to learn mappings between measurable or design variables—such as voltage, current, phase, frequency, coil geometry, and material properties—and target quantities such as transfer efficiency, electromagnetic performance, receiver position, operating state, fault category, and control variables. Once trained, a supervised model can provide computationally efficient inference for previously unseen inputs, although the costs associated with data generation, labeling, and model training must also be considered.

Supervised-learning studies in WPT encompass conventional regression techniques, shallow neural networks, ensemble-learning methods, and deeper learning architectures. In magnetic-coupler design, Ding et al. [75] compared several regression algorithms and selected a random-forest model to predict suitable coupler geometries. Their method reduced a design evaluation requiring approximately 5 min using FEA to a few seconds, corresponding to an approximately 50-fold acceleration. Kim et al. [76] developed a feedforward neural network to predict the quality factor of high-frequency spiral coils from their geometric parameters. For system-wide fault diagnosis, Li and Zhang [77] employed a hierarchical support vector machine (SVM) using measurements from a single current sensor and achieved a fault-identification accuracy exceeding 95%. These studies illustrate the applicability of supervised learning to continuous regression, design-space approximation, and discrete classification.

The application scope of supervised learning has expanded from electromagnetic design and performance prediction to foreign-object detection, fault diagnosis, semiconductor-device modeling, electromagnetic-interference estimation, and automated WPT design. Neural-network and machine-learning models have been employed for foreign-object detection and fault diagnosis [66,77-82], semiconductor-device modeling [83], electromagnetic-interference estimation [84], and magnetic-coupler design [31,75,76,85,87]. Recent studies have also introduced deeper convolutional architectures. Gong et al. [82] developed an enhanced one-dimensional ConvNeXt model for WPT fault classification, whereas Zhang et al. [87] employed a convolutional neural network (CNN) to predict electromagnetic performance and optimize WPT parameters. Generative neural networks and ensemble-learning models have further extended data-driven methods to dynamic inductive-power-transfer design and magnetic-coupler optimization [31,85]. Nevertheless, conventional regression models and deep-learning architectures continue to coexist, with their suitability depending on data availability, task complexity, interpretability requirements, and deployment constraints.

From a task-oriented perspective, supervised learning supports several stages of WPT system development. In structural design, supervised and ensemble models accelerate the evaluation and optimization of coil geometries, ferrite-core configurations, and related design parameters [31,75,76,85,87]. Supervised models have also been employed to predict electromagnetic performance, including the reduction in electromagnetic interference achieved using an intermediate coil [84]. In parameter and state estimation, data-driven models estimate receiver position from transmitter-side electrical measurements [65] and predict coil misalignment from detection-coil and primary-side voltage signals [66]. Supervised learning further supports safety monitoring and diagnostic tasks, including foreign-metal-object detection and system-level fault identification [66,77-82]. In operational control, supervised models have been used for power regulation, efficiency tracking, and maximum-power-point tracking [71,86]. Symbolic regression provides an additional interpretable approach by deriving explicit analytical expressions from data. It has been used to estimate coil separation from simulated S-parameters and experimentally measured Z-parameters and to characterize the frequency-phase relationship of WPT systems, with reported prediction errors below 1% [89].

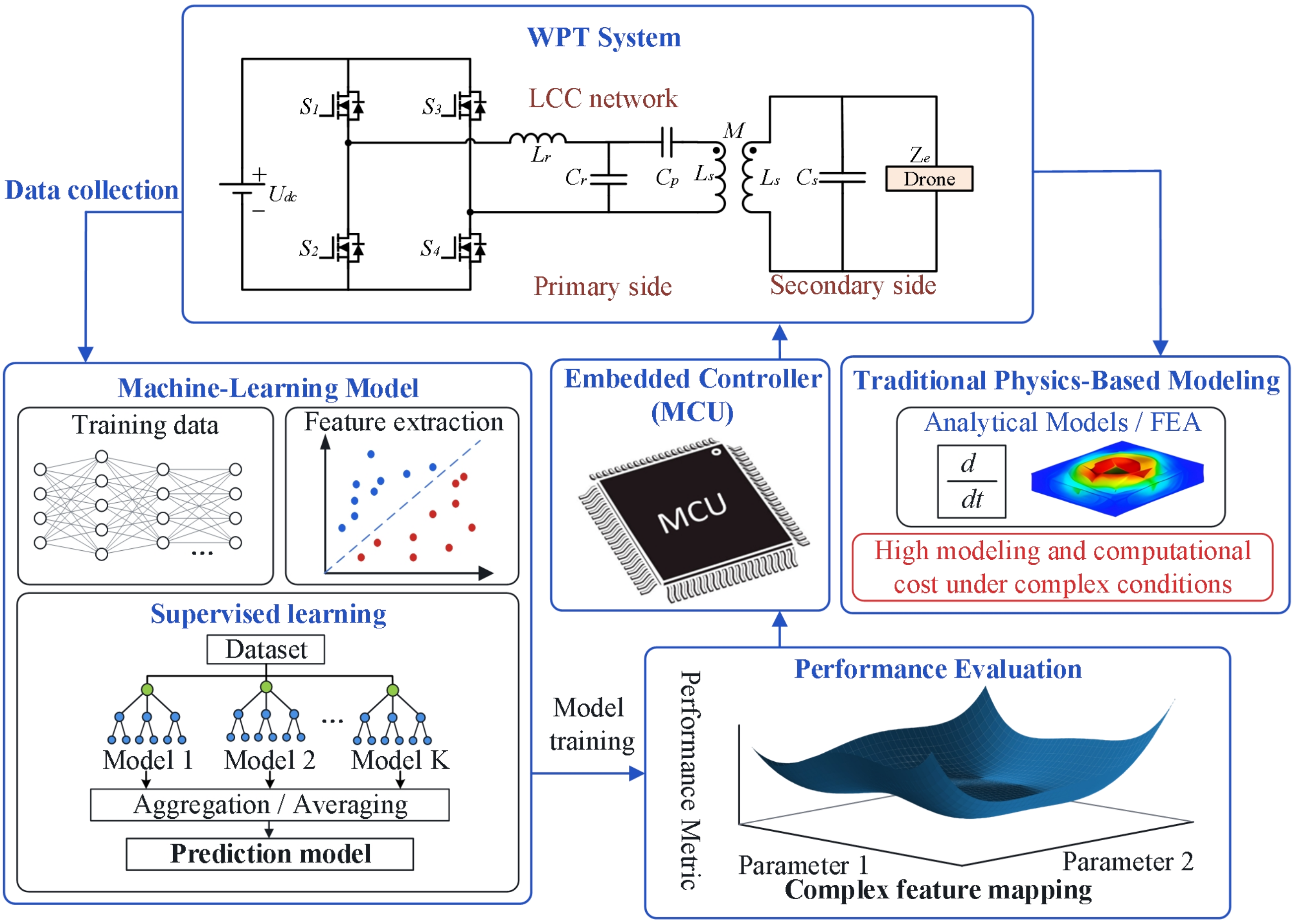

**Figure 4.** Schematic Overview of Typical Supervised Learning Models for WPT System Control under Complex Multi-physics Conditions.

Figure 4 provides a schematic overview of the roles of supervised learning in WPT systems operating under complex conditions. Analytical models offer physical interpretability, whereas FEA provides high-fidelity electromagnetic evaluation at the expense of substantial computational cost when repeatedly applied during optimization. Supervised models offer a complementary approach by approximating nonlinear input–output relationships from labeled data and providing rapid inference after training. Their practical effectiveness, however, depends on the representativeness of the training data, prediction uncertainty, generalization across operating conditions, physical consistency, and computational requirements for deployment. Supervised learning should therefore be viewed as a task-dependent complement to physics-based modeling rather than as a universal replacement for analytical or numerical methods.

### 2) Reinforcement learning

Reinforcement learning (RL) has become increasingly prominent in WPT research, accounting for approximately 22% of the studies included in the reviewed corpus. Unlike supervised learning, RL does not require labeled examples of optimal actions. Instead, an agent learns a decision policy through repeated interaction with an environment. At time step $t$, the agent observes the state $s_t$, selects an action $a_t$, receives a reward $r_t$, and transitions to the next state $s_{t+1}$. The objective is to learn a policy $\pi(a|s)$ that maximizes the expected cumulative discounted reward. In WPT applications, the state may include electrical measurements, coupling conditions, energy states, or wireless-channel information, whereas the action may involve switching frequency, duty ratio, phase shift, compensation capacitance, coil activation, charging trajectory, or resource allocation.

RL methods used in WPT can be broadly divided into value-based, policy-gradient, and actor-critic approaches. Value-based methods, including Q-learning and deep Q-network (DQN), estimate the expected return associated with each state-action pair and are particularly suitable for discrete decision spaces. They have been applied to magnetic-core and coupler optimization [41,50,51], constant-voltage and constant-current regulation [42], intelligent coil-activation decisions under UAV misalignment [72], and charging-path or scheduling problems [23,43]. Policy-gradient methods directly optimize a parameterized policy and are well suited to stochastic or continuous decision problems; proximal policy optimization (PPO), for example, has been investigated for joint UAV trajectory, energy-harvesting schedule, transmission-power, and reconfigurable-intelligent-surface (RIS) phase-shift optimization [95]. Actor-critic methods combine a policy actor with a value-function critic. Representative algorithms include deep deterministic policy gradient (DDPG), twin delayed deep deterministic policy gradient (TD3), soft actor-critic (SAC), and their multi-agent variants. These methods are particularly relevant to high-dimensional continuous control. TD3 reduces the Q-value overestimation and policy instability associated with conventional DDPG and has been employed for offline PI-controller tuning in frequency-controlled LCC-S WPT converters [46] and for Lyapunov-constrained task-offloading and WPT decisions [96]. SAC has also been investigated for continuous power optimization in UAV wireless charging [28].

At the physical-system level, RL has been applied to voltage regulation, output-mode control, frequency tuning, coil activation, and energy-storage management. An early data-driven RL regulator was developed for the output-voltage control of EV wireless chargers [16]. Liu et al. [42] subsequently proposed a dynamic-programming reinforcement-learning method that selects the primary compensation capacitance to maintain constant-voltage or constant-current output under load variations, achieving maximum and average regulation errors of 1.29% and 0.93%, respectively. Wu et al. [72] used Q-learning to determine a minimal activation pathway for a reconfigurable UAV charging array, reducing the number of coil-activation steps by 92.2% relative to exhaustive search. Safari et al. [46] employed TD3 to optimize PI-controller parameters offline for frequency-based voltage regulation, thereby avoiding online policy updates during real-time operation. A comparative study of model-free controllers for WPT-connected hybrid energy storage further showed that DQN control offers rapid transient response, although at a higher computational cost than fuzzy-logic and supervised ANN controllers [155].

RL has also been widely investigated in communication-oriented systems that integrate wireless information transfer, computation, and energy delivery. Representative applications include adaptive computation offloading in WPT-assisted mobile edge computing [91], energy-efficiency optimization in intelligent-reflecting-surface(IRS)-assisted UAV-WPT networks [92], joint UAV trajectory and RIS configuration for age-of-synchronization minimization [93], and user scheduling, power allocation, and power splitting in massive MIMO-NOMA-SWIPT systems [94]. DDPG and PPO have been used to jointly optimize UAV trajectories, transmission power, energy-harvesting schedules, and RIS phase shifts [95], while Lyapunov-optimized TD3 has been developed for joint UAV-assisted task offloading and WPT under time-varying wireless channels [96]. These communication-oriented studies extend RL beyond converter-level control to network-level resource allocation and scheduling, but their system models, state variables, and performance objectives differ substantially from those of electromagnetic and power-electronic WPT control.

The principal advantage of RL lies in its ability to optimize sequential decisions under time-varying and multi-objective operating conditions without requiring labeled optimal actions. Model-free RL can also reduce dependence on an explicit plant model, although it still requires a sufficiently representative training environment and extensive interaction data. Its practical limitations include high training and sampling costs, sensitivity to reward design and hyperparameters, unsafe exploration on physical hardware, limited interpretability and stability guarantees, and performance degradation during simulation-to-reality transfer. Most existing WPT studies therefore train their policies offline in simulation and deploy only the learned policy or optimized controller parameters on the physical system. Safe exploration, constraint-aware learning, hardware-in-the-loop validation, and reliable sim-to-real transfer remain central requirements for engineering deployment.

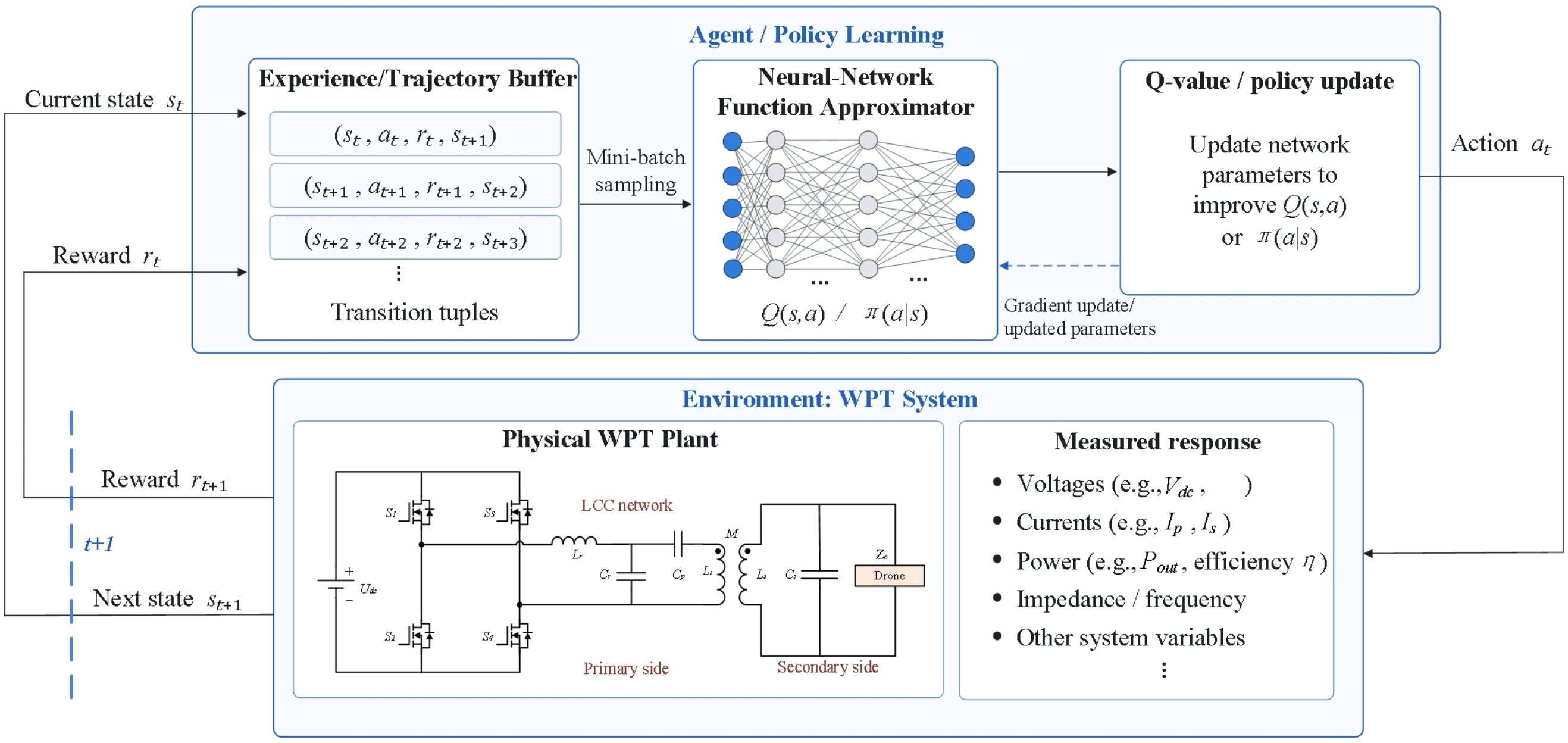

**Figure 5.** Schematic Overview of Typical Reinforcement Learning Models for Adaptive WPT System Control under Dynamic and Complex Operating Conditions.

### 3) Hybrid Intelligent Methods

Hybrid intelligent methods combine data-driven learning with numerical simulation, physical knowledge, metaheuristic optimization, or conventional control to exploit their complementary capabilities. The reviewed studies exhibit four recurring integration patterns: surrogate-assisted optimization, physics-guided learning, learning-assisted conventional control, and multistage offline–online optimization. These patterns are not mutually exclusive, and an individual framework may incorporate elements from more than one category.

The first pattern couples learned surrogate models with numerical simulation or metaheuristic optimization. In biomedical coil design, an artificial neural network (ANN) trained using finite-element-analysis data has been combined with a genetic algorithm and the coyote optimization algorithm to identify geometries that maximize transfer efficiency [55]. For efficiency-optimization control, a multilayer perceptron predicts output power and efficiency, PSO searches for optimal phase-shift combinations offline, and a lightweight ANN maps the resulting optimal settings for real-time implementation [69]. Genetic-algorithm-based symbolic regression provides another form of hybrid model discovery by deriving explicit analytical relationships for coil-distance estimation and frequency–phase characterization [89]. Similarly, field-solver data have been used to train an ANN for compensation-capacitance tuning, with reported maximum errors of 3.40%, 0.88%, and 2.13% for capacitance, efficiency, and output power, respectively [98].

The second pattern integrates physical or mechanism-based knowledge into data-driven models. Physics-informed models can incorporate governing relationships through physically derived training data, explicit constraint layers, auxiliary loss terms, or coupled model components. At the design stage, Pandey et al. pretrained a machine-learning model using analytical-model data and subsequently fine-tuned it using a small FEA dataset, achieving errors below 4% in WPT coil-parameter estimation [57]. This strategy reduced dependence on extensive FEA datasets while retaining the accuracy required for magnetic-coupler modeling. For online parameter and state estimation, impedance-angle, mutual-inductance, and load relationships have been embedded into a neural architecture for the joint estimation of electrical parameters and coil misalignment [61]. In underwater WPT, an ANN pretrained using mechanism-generated data was fine-tuned through transfer learning with a limited amount of experimental data, enabling joint identification of mutual inductance and load with prediction errors below 3% and a response time within 2.72 ms [62]. These approaches seek to improve data efficiency and physical consistency without relying exclusively on either analytical models or experimental data.

The third pattern integrates learning models with established control structures. An extreme learning machine has been employed to estimate uncertainty bounds online within a fixed-time sliding-mode voltage controller, reducing settling time by 45.70%–51.72% relative to the investigated PI-control baselines [90]. In another framework, XGBoost surrogate models predict the dynamic performance of candidate adaptive proportional–integral controllers, while an improved PSO algorithm determines controller-gain lookup tables for real-time implementation [10]. These approaches retain the stability structure and engineering familiarity of conventional controllers while using learning and optimization to improve adaptation or controller design.

The fourth pattern coordinates offline optimization, surrogate modeling, and online adaptation within a multistage framework. A unified EV-WPT framework combines genetic-algorithm-based static optimization, ANN surrogate modeling, and proximal-policy-optimization-based dynamic control to address both nominal efficiency and operating-condition variations [11]. In SWIPT-enabled wireless sensor networks, genetic algorithms, ant colony optimization, and proximal policy optimization have been assigned to cluster-head deployment, cluster formation, and dynamic resource allocation, respectively [99]. For shared autonomous vehicles supplied by dynamic wireless charging, deep reinforcement learning has been integrated with a routing heuristic to jointly reduce route cost and stabilize the vehicle state of charge, thereby supporting continuous vehicle operation [100]. Offline–online parameter-estimation schemes similarly combine initial model training with online model correction to reduce data requirements and accommodate operating variations [63].

These studies indicate that hybrid intelligent methods are evolving from simple algorithm combinations toward coordinated workflows spanning modeling, optimization, estimation, and control. Nevertheless, most existing architectures remain application-specific, and their reported performance is evaluated using different datasets, objectives, and operating conditions. Further research is therefore required to establish standardized comparisons, quantify the contribution of individual components through ablation studies, provide stability and safety guarantees, and validate computational feasibility on practical embedded hardware. Figure 6 illustrates the representative pathways through which data-driven models are integrated with physical knowledge, metaheuristic optimization, conventional control, and reinforcement learning in WPT systems.

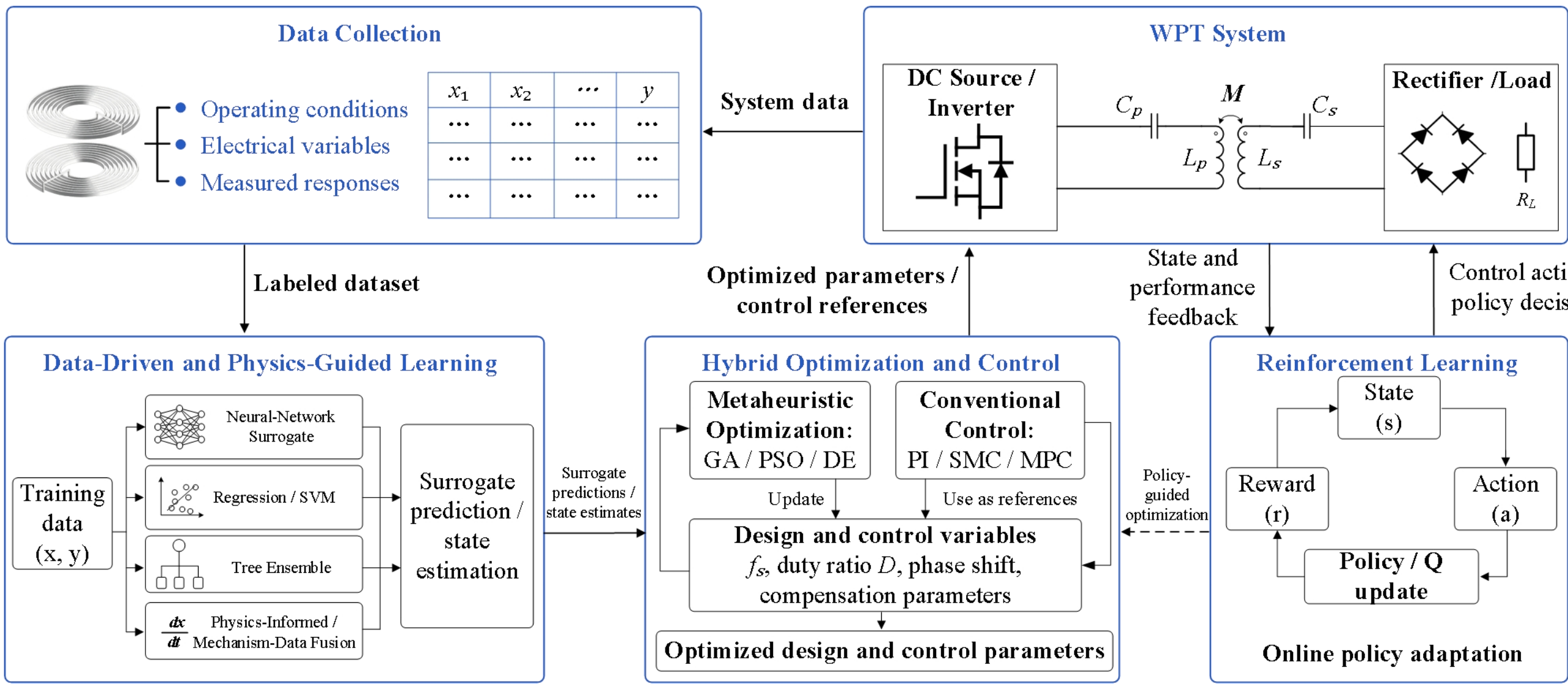

**Figure 6.** Representative integration patterns of hybrid intelligent methods in WPT systems.

## 3.2 Algorithm Comparison

The preceding subsections reviewed the methodological characteristics and representative WPT applications of supervised learning, reinforcement learning, and hybrid intelligent methods. Selecting an appropriate approach for a specific engineering task, however, requires consideration of the available data, physical-model assumptions, computational requirements, deployment conditions, and decision structure. For example, position-adaptive control in dynamic charging may combine supervised position estimation with a conventional controller, directly learn a control policy through reinforcement learning, or integrate physical models with data-driven estimation and control. The following comparison clarifies the principal trade-offs and conditions of applicability of these three methodological categories.

### 1) Data dependency and information sources

Supervised learning requires labeled input–output pairs obtained from FEA, analytical calculations, or experimental measurements. The amount of data required depends on task complexity, operating-range coverage, model capacity, and the desired level of generalization. Simulation-generated labels can reduce experimental effort but may introduce simulation-to-experiment bias, whereas experimentally acquired labels are generally more representative but costly to collect and annotate.

Reinforcement learning does not require target labels. Instead, it learns from interaction trajectories comprising states or observations, actions, rewards, and subsequent states. Simulation environments enable safer and more extensive exploration than direct training on physical hardware, but generating sufficient interaction data can still be computationally expensive, and discrepancies between simulated and physical environments may produce a sim-to-real gap. Hybrid methods have more variable data requirements. Physics-informed constraints, mechanism-generated data, transfer learning, and simulation-based pretraining may reduce the amount of experimental data required, although the resulting performance depends on the validity of the incorporated physical assumptions.

Supervised models and model-free reinforcement-learning agents do not necessarily require an explicit plant model during online inference, although their training data or environments may be generated using analytical or numerical models. Model-based reinforcement learning explicitly incorporates an environment model. Hybrid methods may combine physical models, learned surrogates, metaheuristic optimizers, or conventional controllers. Their information sources may therefore include measured data, simulation outputs, governing equations, expert knowledge, optimization feedback, and control signals.

### 2) Accuracy and Computational Characteristics

The performance of supervised learning depends on training-data coverage, label quality, feature selection, and model capacity. Supervised models can achieve high regression or classification accuracy within the operating conditions represented by the training data, as demonstrated in foreign-object detection, system-level fault diagnosis, and semiconductor-device modeling [77-79,83]. Reinforcement learning is evaluated primarily through closed-loop criteria, including tracking error, transfer efficiency, cumulative reward, and constraint satisfaction, and its performance depends strongly on the state and action representations, reward formulation, and training environment [16,42]. Hybrid methods incorporate physical models, electromagnetic constraints, or established control structures to improve data efficiency and physical consistency, although their performance remains dependent on the validity of the embedded assumptions [61,62,90].

The principal computational advantage of learning-based methods is that expensive electromagnetic simulations, numerical inversion, or policy optimization can be transferred to the offline stage. Once trained, surrogate models, state estimators, and learned control policies can provide rapid inference during deployment [14,16,42,62,65,69,75,90]. This offline-to-online transfer of computational effort facilitates iterative structural design, online state estimation, and real-time control, although practical performance still depends on model complexity, deployment hardware, and the required sampling or control period.

### 3) Typical application scenarios and optimization objectives

Supervised learning has been applied throughout WPT system development. It supports surrogate modeling for magnetic-coupler design, regression and classification for parameter and state estimation, foreign-object and fault detection, and learned mappings for operating-point or control-variable prediction [65,66,68,69,71,75,76]. Reinforcement learning is particularly suitable for sequential decision problems, including voltage and frequency regulation, transmitter or beam selection, charging-path planning, energy scheduling, and communication–energy resource allocation [16,23,42,91-96]. Hybrid methods encompass surrogate-assisted structural optimization, physics-guided parameter identification, learning-assisted conventional control, and multistage or cross-layer optimization [10,11,55,61-63,69,90,99].

The three categories also differ in their relationship to optimization. Supervised learning primarily constructs predictive or classification models. When applied to design optimization, a learned surrogate is generally coupled with a gradient-based or derivative-free optimizer. Reinforcement learning directly learns policies for sequential decisions and represents long-term objectives through cumulative rewards, although its multi-objective performance depends on reward formulation and constraint handling. Hybrid frameworks can coordinate prediction, optimization, estimation, and control across discrete and continuous variables or different time scales, but their additional components increase implementation complexity and make it more difficult to attribute performance improvements to individual components.

Several conclusions emerge from this comparison. First, no method is universally superior; method selection should reflect data availability, physical-model fidelity, online latency, safety requirements, and the temporal structure of the task. Second, the three categories are complementary and can be combined across design, estimation, and operational control, as demonstrated by multistage surrogate–optimization–control frameworks [11,69,99]. Third, the boundaries among these categories are becoming less distinct as physics-guided learning, transfer learning, generative models, and offline–online adaptation are incorporated into WPT workflows [61-63,85]. Establishing common benchmarks and reporting standards will therefore be essential for determining whether increased algorithmic complexity produces meaningful improvements in engineering performance.

Table 3 consolidates the preceding comparison in terms of data requirements, model dependence, information sources, computational burden, online response, performance characteristics, representative WPT tasks, and decision structure.

**Table 3.** Comparison of machine-learning categories for WPT applications

| Dimension | Supervised learning | Reinforcement learning | Hybrid methods |
|---|---|---|---|
| Data dependency | Labeled simulation or experimental data | State–action–reward trajectories | Data combined with physical priors |
| Model dependence | Low explicit dependence; data may embed model assumptions | Model-free or model-based; environment-dependent | Explicit integration of physical models or constraints |
| Information sources | Measurements, simulations, and analytical data | States, actions, rewards, and transitions | Data, physical knowledge, and engineering constraints |
| Computational efficiency | Costly offline training; efficient inference | High training cost; efficient policy inference | Dependent on model coupling and implementation |
| Accuracy characteristics | High in-distribution accuracy; limited extrapolation | Task-specific closed-loop performance | Improved physical consistency; model-dependent |
| Response speed | Rapid inference after training | Rapid policy execution after training | Rapid for lightweight models; slower for iterative schemes |
| Typical applications | Design, prediction, estimation, diagnosis, and control | Adaptive control, coordination, and scheduling | Physics-informed modeling, optimization, identification, and control |
| Decision structure | Predictive/classification mapping; coupled with optimizer for design | Sequential decision-making; policy optimization | Coordinated prediction, optimization, estimation, control; multi-stage/multi-timescale |

## 4. Specific applications of machine learning in wireless power transfer

The performance of WPT systems depends on the coordinated treatment of three interdependent task domains: structural design and optimization, parameter and state estimation, and operational control. Conventional approaches provide established engineering solutions but may involve computationally intensive electromagnetic simulations, additional sensing or modeling requirements, and limited adaptability to changing operating conditions. Machine learning provides complementary tools for addressing these limitations. Surrogate models and reinforcement-learning-based search can accelerate design-space exploration; supervised and hybrid models can infer system states from measurable responses; and learned mappings or reinforcement-learning policies can support adaptive regulation and scheduling. Accordingly, this section reviews machine-learning-enabled WPT research across these three task domains and compares the roles of supervised learning, reinforcement learning, and hybrid methods.

Despite recent progress, the transition from laboratory demonstrations to practical WPT systems remains constrained by data scarcity and bias, limited physical consistency, the safety of online adaptation, embedded-deployment constraints, and insufficient explainability. Table 4 summarizes these challenges and the corresponding research directions.

**Table 4.** Key challenges and potential mitigation strategies for ML-enabled WPT

| Challenge | Implications for WPT | Potential mitigation strategies |
|---|---|---|
| Data scarcity and bias | Experimental, fault, and high-power transient data are costly, incomplete, and potentially imbalanced. | Hybrid simulation–experimental datasets, active learning, data augmentation, uncertainty quantification, and shared benchmarks |
| Limited physical consistency | Data-driven predictions may violate resonance, thermal, safety, or device constraints. | Physics-informed features and loss functions, constrained learning, hybrid modeling, and cross-condition validation |
| Safe online adaptation | Online exploration may cause overcurrent, overvoltage, hard switching, or overheating. | Offline or digital-twin training, safe reinforcement learning, constrained action spaces, safety filters, and supervisory fallback |
| Embedded-deployment constraints | Real-time WPT applications impose strict limits on computation time, memory, and power consumption. | Model compression, pruning, quantization, knowledge distillation, and DSP/FPGA implementation |
| Explainability and maintenance | Practical systems require auditable decisions, uncertainty awareness, and detection of faults or model drift. | Interpretable models, feature attribution, confidence estimation, residual monitoring, and drift detection |

## 4.1 Structural Optimization

Transfer performance is strongly influenced by structural and material parameters, including the number of coil turns, coil geometry, coil dimensions, air gap, and magnetic-core properties. Conventional magnetic-coupler design commonly combines FEA with parameter sweeps or numerical optimization. Although parameter sweeps are straightforward, their computational cost increases rapidly with the number of design variables, while population-based optimizers may require numerous FEA evaluations and can exhibit premature convergence. Learning-based methods address these limitations by constructing surrogate mappings between design variables and performance metrics or by learning search policies for design-space exploration. The existing approaches can be broadly divided into supervised surrogate modeling, reinforcement-learning-based search, and hybrid learning–optimization methods.

### 1) Supervised Learning as an FEA Surrogate

After offline training, supervised models can replace repeated FEA evaluations with rapid predictions of electromagnetic performance. Wen et al. employed a BPNN to predict the optimal receiver-coil radius [26], obtaining a coefficient of determination of ($R^2=0.9977$); the predicted optimum differed from the FEA result by 5.43%. Mitsuhashi et al. used a DNN to optimize a thin-coil configuration [29]. For the reported design conditions, the optimized configuration increased the quality factor by 49% and improved the power-transfer efficiency from 85.4% to 90.0%. Xie et al. incorporated low-fidelity analytical estimates derived from the mirror method as prior inputs to a neural network [27]. Using 500 FEA samples, the proposed model achieved mean errors below 3% for mutual-inductance and loss prediction, although the maximum error depended on the coupler structure and predicted quantity. Kim et al. developed a feedforward neural-network-based method for spiral-coil design [30], reporting a quality-factor error below 4% and an intersection-over-union value above 78% with fewer than 1,000 training samples. Ding et al. proposed a stacking ensemble surrogate model for magnetic-coupler design [31], reducing the evaluation time from several minutes for FEA to several seconds and achieving an approximately 50-fold acceleration.

Neural-network-based surrogate models have also been used to predict the mutual inductance of multiphase and three-dimensional couplers [32,37], optimize DD-coil and ferrite-core geometries [33], design passive shielding structures and active compensation coils for magnetic-field mitigation [34,35], and support the design of underwater and implantable WPT systems [19,36]. In the latter applications, the learned surrogate is coupled with a multi-objective optimizer; the complete optimization workflows are therefore further discussed under hybrid methods.

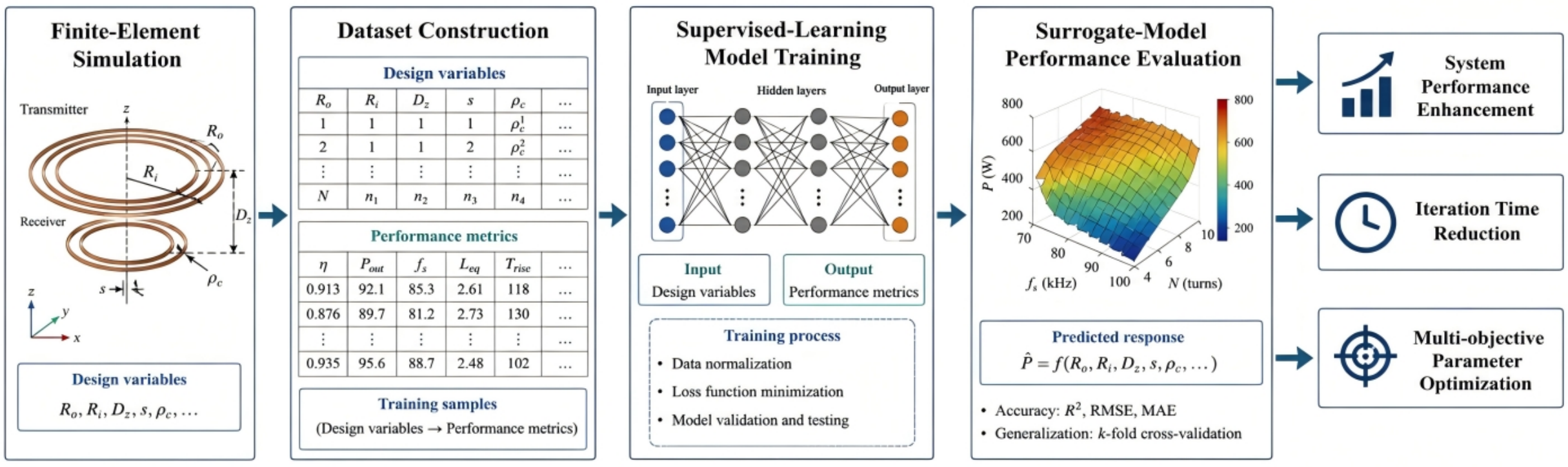

**Figure 7.** Supervised Learning Agent Model Workflow Diagram.

### 2) Reinforcement Learning for Design-Space Exploration

Reinforcement learning formulates structural design as a Markov decision process, in which an agent explores the design space through interaction with a simulation environment and learns a policy according to task-specific rewards. Choi et al. applied RL to ferromagnetic-core structural design [50]. Using 300 training episodes, equivalent to approximately 2.3% of the 12,870 possible structures, the method identified a core configuration whose experimentally measured coupling coefficient was 7% higher than that of the reference design at a coil separation of 98 cm. Cong et al. subsequently developed an improved deep-reinforcement-learning framework for DD-type magnetic-coupler design [41]. Compared with a conventional fully filled core, the optimized receiver-side structure reduced the core volume by 28% while increasing the mutual inductance by 2.3%.

Rahulkumar and co-workers employed epsilon-greedy Q-learning to optimize ferrite-core placement in different coupler geometries [51,52]. Their methods selected 21 ferrite-core positions from 104 candidate positions while considering coupling performance, core thickness, and multilayer arrangements. For the double-D-and-O coupler, the optimized configuration reduced the coupler weight by more than 50% relative to the unoptimized design and also reduced material requirements and cost [51].

Beyond magnetic-coupler design, RL has been applied to broader infrastructure-level optimization. Representative studies include cost-aware design of wireless-charging electric-bus systems [22], joint placement of static and dynamic EV charging infrastructure [18], deployment of multiple UAV chargers for IoT devices [44], and optimal charger placement in wireless rechargeable sensor networks [45]. These applications demonstrate the ability of RL to address discrete, sequential, and system-level design decisions, although their objectives and constraints differ substantially from those of electromagnetic component optimization.

### 3) Hybrid Method

Hybrid methods integrate learning models with metaheuristic optimization, iterative data refinement, or adaptive control to combine the complementary strengths of different techniques. Based on the reviewed studies, these methods can be broadly organized into three representative patterns: learning-assisted metaheuristic optimization, multi-model optimization with iterative data refinement, and offline design optimization combined with online adaptive control.

Thiagarajan et al. combined Taylor-based Firefly and Dove Swarm Optimization with a Bessel-

filter-enhanced fuzzy recurrent neural-network controller [24]. The experimentally validated system achieved a peak transfer efficiency of 95% at an air gap of 135 mm. Bennia et al. developed a neural-network surrogate with a reported prediction accuracy above 97% and coupled it with genetic and coyote optimization algorithms [55], obtaining an optimized transfer efficiency of up to 76% for implantable WPT coils. Fawad et al. integrated a hybrid residual-layer sequential neural network, support vector regression, and a deep Q-network with K-means-based iterative sample refinement [53]. Compared with ANSYS Maxwell simulations, the reported acceleration factors were approximately 4,300, 16,960, and 10,138 for the three respective models.

Additional studies have extended hybrid surrogate-assisted optimization to application-specific coupler design. Alotaibi combined a hybrid sequential neural network with multivariate regression to optimize the coil winding and ferrite-core geometry of a mobile-device WPT system. The optimized solid-core configuration achieved a mutual inductance of 12.52 μH and was validated through FEA and experiments [88]. Xie et al. developed an electromagnetics-embedded BPNN for the multi-objective design of an undersea planar magnetic coupler. Relative to a purely data-driven model, the proposed method required less than half the training data and reduced the optimization time by at least six orders of magnitude; FEA and 1.5-kW prototype results agreed with the model predictions within 5.1% [105]. Zhang et al. combined a DNN surrogate with dynamic Lévy-flight Beluga Whale Optimization for multi-objective resonant-coil design, reporting FEA-relative errors of 0.2% for self-inductance and 1.3% for mutual inductance, together with a convergence time of 7.8 s [109]. Wu et al. combined a deep neural-network surrogate with a self-adaptive discrete PSO algorithm [54]. The optimization was completed in less than 20 s and validated using a 2 kW LCC-S WPT platform; at a test power of 1 kW and under maximum misalignment, the optimized coil achieved a transfer efficiency of 95.02%. Ben Fadhel et al. further proposed a cross-stage framework integrating offline genetic-algorithm optimization, ANN surrogate modeling, and online reinforcement-learning control based on proximal policy optimization [11]. These studies demonstrate that hybrid methods can coordinate surrogate prediction, design-space search, and adaptive control, although their additional algorithmic components increase implementation and validation complexity.

## 4.2 Parameter and State Estimation

Key WPT parameters and operating states, including mutual inductance, load resistance, coupling conditions, and receiver position, vary with coil misalignment, load changes, temperature, and the surrounding medium. Conventional estimation methods generally rely on analytical circuit models, auxiliary sensing devices, lookup tables, or recursive estimators. Although these approaches are physically interpretable, their accuracy may deteriorate because of model mismatch, measurement noise, or operating conditions not considered during calibration. Machine-learning methods provide a complementary approach by learning nonlinear mappings from measurable electrical responses to the quantities of interest. Existing studies can be broadly grouped into direct data-driven parameter estimation, hybrid model- and data-driven estimation, receiver-position and misalignment estimation.

### 1) Direct Data-Driven Parameter Estimation

Supervised learning can directly estimate parameters such as mutual inductance, load resistance, and coupling coefficient from transmitter-side voltage, current, phase, impedance, or harmonic-domain features. He et al. simultaneously estimated mutual inductance and load resistance from primary-side information, reporting maximum errors of 2.93% and 7.4%, respectively [56]. Kim et al. used transmitter-side voltage and current features to predict the load condition and coupling coefficient without requiring receiver-side sensing or communication [58]. Yang et al. extracted harmonic-domain features using the fast Fourier transform and reported relative mutual-inductance estimation errors below approximately 1.2% for most evaluated cases [59]. Wang et al. developed a k-nearest-neighbor-based method for mutual-inductance identification in underwater WPT systems and compared its performance with SVM, random forest, XGBoost, and BPNN models [60]. These studies demonstrate that measurable electrical responses can support parameter estimation without direct access to the receiver, although the resulting accuracy remains dependent on data coverage and measurement quality.

### 2) Hybrid Model- and Data-Driven Estimation

Hybrid estimation methods incorporate analytical relationships, mechanism-generated data, or physical constraints into the learning process. Tong et al. embedded the physical relationships of a series-series compensated WPT system into a physics-informed neural network. Experiments on a 10 kW platform yielded relative errors of 7.7%, 2.1%, 2.0%, and 1.8% for coil displacement, mutual inductance, leakage inductance, and load resistance, respectively [61]. Chen et al. combined mechanism-generated data with experimental adaptation for an underwater WPT system, obtaining prediction errors below 3% under three representative underwater conditions and a response time within 2.72 ms [62]. Cheng et al. combined an online magnetic positioning system with neural-network models to estimate the coupling coefficient and secondary-side voltage and current from auxiliary-coil signals and primary-side measurements. Experiments on a 500-W prototype demonstrated communication-free parameter estimation and feedback control under variable coil positions [115]. These results indicate that physical knowledge can reduce reliance on extensive experimental datasets and improve consistency with established system relationships. Nevertheless, the effectiveness of such approaches remains dependent on the validity of the embedded model and the representativeness of the calibration data.

### 3) Receiver Position and Misalignment Estimation

Receiver position and coil misalignment can be inferred from electrical responses or auxiliary sensing information. Dai et al. employed a long short-term memory network for receiver localization in a multi-transmitter single-receiver WPT system and reported an average position error below 2.40 mm [49]. Yan et al. used a BPNN to predict the receiver position in a dynamic WPT system, obtaining a test-set correlation coefficient of approximately 0.9994 [25]. Shen et al. compared BPNN and support vector regression models for receiver-position estimation, reporting an average relative error of approximately 5%. Based on the estimated position, an adaptive transmitter-switching strategy maintained system efficiency close to 90% under misalignment conditions [65]. Özüpak and Aslan trained an ANN using FEM-generated data to predict the coupling coefficient, mutual inductance, and transfer efficiency under aligned and laterally misaligned coil configurations [38]. More recently, Li et al. developed a voltage-fingerprint-based position-identification and closed-loop alignment method using a residual-calibrated ensemble network. Experiments conducted over a two-dimensional workspace yielded a root-mean-square positioning error of 1.83 cm under electromagnetic interference and measurement noise [113]. Lin et al. combined magnetic-resonance-based localization with reinforcement-learning-based charging-path scheduling, achieving a positioning error of approximately 3.5 cm and an average charging-efficiency improvement of 20.9% [39].

Machine learning has also been used to process position-related and safety-monitoring information jointly. Gong et al. developed a neural-network method for both coil-misalignment prediction and metallic-foreign-object detection [66]. Foreign-object detection should nevertheless be regarded as a related safety-monitoring task rather than as a WPT state variable. Bai et al. compared random forest and DNN models for receiver-position classification, reporting accuracies of 94.05% and 95.99%, respectively, for a 16-class case, together with an average power-transfer-efficiency prediction error below 1% [67]. Bertoluzzo et al. combined camera images with a CNN to estimate mutual inductance in a dynamic WPT system [68]. Symbolic regression provides an additional interpretable alternative by deriving explicit analytical relationships between coil separation and measured S- or Z-parameters, although the reported accuracy varies with the selected parameter representation [64].

The reviewed studies show that learning-based estimators can provide rapid parameter and state inference after offline training. Their practical reliability, however, depends on measurement quality, training-data coverage, uncertainty quantification, and generalization across different alignments, loads, temperatures, transmission media, and system topologies.

## 4.3 Control and Optimization

Machine-learning methods have been investigated for constant-current and constant-voltage regulation, impedance matching, frequency tuning, efficiency optimization, transmitter selection, and energy scheduling. Conventional PI and PID controllers remain attractive because of their simple implementation and low online computational burden, but fixed controller parameters may require retuning when coupling conditions, loads, or system dynamics vary substantially. Learning-based control provides complementary mechanisms for approximating control mappings, optimizing operating parameters, and learning sequential control policies. Existing studies can be grouped into supervised-learning-based control, reinforcement-learning-based control, and hybrid control.

### 1) Supervised-Learning-Based Control

Supervised control methods learn mappings from measured operating conditions to control variables or desired operating points. Jeong et al. developed a neural-network-based method for real-time impedance matching and maintained a power-transfer efficiency of approximately 90% over a transmission-distance range of 10-25 cm [14]. Wang et al. employed a multi-cascade CNN to predict compensation-capacitance settings and achieved lower prediction errors than a conventional network structure [70]. Na et al. trained a machine-learning model using experimentally measured dc voltage and current data to support rapid transmitter-side power control under intermittent receiver feedback [71]. Das et al. employed a multilayer long short-term memory network to predict the lateral vehicle position associated with the maximum electromagnetic-field strength and to adapt vehicle motion during dynamic wireless charging. In simulation, the proposed method achieved up to 174% higher charging efficiency than lane-center motion-control baselines [97]. These approaches avoid online exploration and can provide rapid inference after training, but their performance is constrained by the operating conditions represented in the training dataset.

### 2) Reinforcement-Learning-Based Control

Reinforcement learning formulates WPT control as a sequential decision-making problem in which an agent selects control actions according to measured states and reward feedback. Teng et al. employed Q-learning for data-driven voltage regulation and demonstrated recovery from a load disturbance within approximately 4 ms in the reported simulation study [16]. Liu et al. developed a dynamic-programming-based reinforcement-learning strategy for constant-current and constant-voltage regulation, reporting maximum and average output errors of 1.29% and 0.93%, respectively, with negligible influence on system efficiency [42].

Reinforcement learning has also been applied to frequency control, transmitter-coil activation, UAV charging, and energy scheduling. Dai et al. developed a soft-actor-critic-based power-optimization method for UAV wireless charging [28]. Wu et al. used deep reinforcement learning to identify a minimal transmitter-coil activation pathway, reducing the number of activation steps by 92.2% relative to exhaustive search on an 8 × 8 transmitter array, with approximately five coil activations required on average [72]. Safari et al. employed twin delayed deep deterministic policy gradient to optimize PI-controller parameters offline for frequency regulation in an LCC-S resonant converter [46]. Qing et al. combined clustering, adaptive frequency regulation, energy routing, and shielding strategies to improve system efficiency from 70.43% to 86.20% under severe misalignment conditions; the reported improvement therefore reflects the complete framework rather than Q-learning alone [40].

For networked and system-level applications, Lee employed deep deterministic policy gradient for adaptive WPT beam scheduling for non-static Internet-of-Things devices [47]. Li et al. used proximal policy optimization for a lunar multi-energy WPT system, reducing energy-supply interruptions by 54.6% and improving beam-alignment efficiency by 41.3% relative to static scheduling [48]. Lu et al. applied a dueling deep Q-network to multimodal UAV charging and logistics scheduling [43], while Ouamri et al. used deep deterministic policy gradient for the joint optimization of energy efficiency, throughput, UAV trajectory, and resource allocation [74]. Omeke et al. investigated Q-learning and state-action-reward-state-action learning for AUV trajectory optimization in an underwater simultaneous wireless information and power transfer network. The proposed approach achieved an energy-efficiency improvement of up to 207% relative to a random-trajectory baseline [73].

Despite their adaptability, reinforcement-learning methods remain subject to substantial offline training costs, sensitivity to state and reward definitions, exploration-related safety risks, and sim-to-real transfer challenges.

### 3) Hybrid and Surrogate-Assisted Control

Hybrid control combines supervised surrogate models, optimization algorithms, reinforcement learning, or established controller structures. Ben Fadhel et al. integrated genetic-algorithm-based static optimization, ANN surrogate modeling, and proximal-policy-optimization-based dynamic control within a unified digital-twin framework [11]. Xu et al. combined a multilayer perceptron with PSO for output prediction and efficiency optimization. Experimental results obtained using a 430 W/85 kHz prototype showed a mean absolute percentage error below 3.5% for the predictive model and a maximum system efficiency of 93.5% [69].

These hybrid approaches combine rapid data-driven prediction with optimization or adaptive decision-making and can reduce repeated evaluations of computationally demanding system models. Their implementation and validation requirements, however, are generally greater than those of a single learning model or conventional controller.

#### 4) Comparative Assessment

The principal computational advantage of learning-based control lies in shifting expensive model evaluation, parameter optimization, or policy learning to the offline stage and providing rapid inference during deployment. This does not imply that learning-based controllers are universally faster than conventional controllers. PI and PID controllers typically have a low and predictable online computational burden. Supervised controllers can provide rapid decisions without online exploration but depend strongly on training-data coverage. Reinforcement-learning policies support sequential and long-horizon decision-making but require extensive training and carefully designed rewards. Hybrid approaches combine data-driven prediction with optimization or physical knowledge, although this additional flexibility introduces greater implementation and validation complexity.

Representative differences among these approaches are summarized qualitatively in Table 5. Numerical results reported in individual studies should not be interpreted as standardized cross-platform rankings because the evaluated tasks, operating ranges, hardware platforms, and performance metrics differ substantially. Conventional PI/PID and fuzzy-control methods included in the table have been reviewed in Section 2.2.

**Table 5.** Reported performance characteristics of representative control methods for WPT.

| Approach | Typical WPT roles | Online computation | Adaptability | Main limitations |
|---|---|---|---|---|
| PI/PID control | Voltage, current, and frequency regulation | Low | Limited outside the tuned operating region | May require retuning under substantial parameter variations |
| Fuzzy control | Nonlinear regulation and operating-mode adjustment | Low to moderate | Moderate | Rule design and tuning depend on expert knowledge |
| Supervised learning | Impedance matching, compensation prediction, and power-control mapping | Low after training | Dependent on training-data coverage | Limited generalization under distribution shift |
| Reinforcement learning | Voltage regulation, frequency tuning, transmitter selection, beam scheduling, and energy management | Low to moderate after training; high during training | Moderate to high within the trained operating domain | Training cost, reward design, exploration safety, and sim-to-real transfer |
| Hybrid control | Surrogate-assisted | Implementation- | Potentially high | Increased |
| | optimization and offline-online adaptive regulation | dependent | | implementation and validation complexity |

## 5. Challenges and Future Directions

The preceding sections reviewed supervised learning, reinforcement learning, and hybrid intelligent methods across three central WPT tasks: structural design and optimization, parameter and state estimation, and control and optimization. Learning-based methods have demonstrated considerable potential to reduce repeated electromagnetic simulations, infer operating quantities from measurable responses, and support adaptive decision-making. Nevertheless, their transition from laboratory studies to practical WPT systems remains constrained by data scarcity, distribution shift, limited physical consistency, embedded-resource constraints, safety requirements, and the absence of standardized evaluation protocols. This section summarizes the common limitations of the existing literature and discusses six future research directions: data-efficient and transferable learning, physics-informed hybrid modeling, lightweight edge deployment, multi-agent coordination, standardized benchmarks, and digital-twin-assisted development.

### 5.1 Common Limitations of Current Research

High-quality labeled data remain difficult and costly to obtain for WPT applications. Simulation data can be generated systematically and can cover broad parameter ranges, but differences between numerical models and physical hardware inevitably introduce a simulation-to-reality gap. Experimental data more directly represent practical operating conditions, but their acquisition is time-consuming and usually covers only a limited range of configurations. Data associated with faults, component degradation, foreign objects, and other abnormal conditions are particularly scarce. Consequently, models trained using limited or imbalanced datasets may not generalize reliably to safety-critical operating conditions.

Many supervised surrogate models rely on datasets generated through large numbers of finite-element simulations. When the coupler topology, operating frequency, material properties, or transmission medium changes, additional sampling and model adaptation are often required. Several hybrid approaches have been developed to alleviate this limitation. Chen et al. pretrained a model using mechanism-generated data and subsequently adapted it using a limited amount of experimental data [62]. Cheng et al. first constructed an approximate BPNN using a small offline dataset and then updated the model online using primary-side measurements [63]. Xie et al. incorporated a low-fidelity analytical model based on the mirror method as prior input to a neural network for underwater magnetic-coupler modeling [27]. Although these approaches reduce exclusive dependence on high-cost experimental or finite-element data, their effectiveness still depends on the validity of the physical model and the representativeness of the adaptation data.

The generalization capability of existing models also remains insufficiently characterized. Most studies evaluate their methods using a specific topology, coupler configuration, frequency range, load range, or experimental platform. Data-driven models may provide high accuracy within the operating region represented by the training dataset, but their errors can increase substantially under out-of-distribution conditions. Practical WPT systems are subject to simultaneous variations in component parameters, coil alignment, load conditions, temperature, transmission media, and measurement noise. Future evaluations should therefore distinguish among simulation-based validation, laboratory prototype validation, hardware-in-the-loop testing, and field deployment. The operating boundaries within which a model maintains acceptable accuracy and stability should also be reported explicitly.

The trade-off between prediction accuracy and deployment cost constitutes another engineering constraint. Deep models can represent complex nonlinear relationships but may require substantial memory, computational resources, and inference time. Lightweight models are easier to deploy but may provide insufficient representational capacity for high-dimensional or strongly nonlinear tasks. Moreover, the timing requirements of WPT monitoring and control vary considerably across functions, ranging from relatively slow supervisory scheduling to fast regulation and protection loops. Computational performance should therefore be evaluated against the timing requirements of the specific task rather than described using a universal real-time threshold.

Existing studies have begun to shift computationally intensive processes offline. For example, optimized adaptive PI-controller gains have been stored in interpolation-capable lookup tables for real-time implementation [10]. Symbolic regression has also been used to derive explicit analytical expressions for distance estimation and frequency-related WPT characteristics [89]. These approaches can reduce online computation, but systematic methods are still required to balance model accuracy, inference latency, memory consumption, and hardware cost.

Physical consistency and interpretability present an additional challenge. Purely data-driven models may generate predictions or control actions that violate energy conservation, circuit relationships, electromagnetic exposure limits, device ratings, soft-switching conditions, or other engineering constraints. Physics-informed neural networks address this issue by incorporating physical relationships into the learning objective or model structure [61]. Symbolic regression derives explicit analytical expressions from data and can therefore provide more transparent input-output relationships [64]. Prior-knowledge-input models incorporate approximate analytical calculations as additional model inputs [27]. These approaches improve physical consistency or transparency, but they do not automatically guarantee stability, safety, or generalization beyond the conditions represented during model development.

## 5.2 Future Research Directions

### 1) Data-Efficient and Transferable Learning

Because experimental measurements and high-fidelity simulations are costly, future research should improve the transferability and data efficiency of learning models. Transfer learning, domain adaptation, few-shot learning, and adaptive sampling could reduce the amount of additional data required when the coupler geometry, topology, or operating range changes. Fawad et al. employed K-means clustering and iterative data refinement to select representative design points and update the training dataset using electromagnetic simulation feedback [53]. Further work should investigate how such strategies can reduce data-generation costs while maintaining accuracy across different WPT systems.

### 2) Physics-Informed and Hybrid Learning

Integrating physical knowledge with machine learning is a promising route toward improved data efficiency, generalization, and physical consistency. Circuit relationships, electromagnetic constraints, and boundary conditions can be incorporated through model inputs, architectural constraints, or physics-based loss functions [61]. Hybrid frameworks can also combine mechanism-generated data with limited experimental measurements, as demonstrated in underwater parameter identification [62]. Future studies should extend these approaches to structural optimization and closed-loop control while explicitly considering electrical, thermal, electromagnetic, and hardware safety constraints.

### 3) Lightweight Models and Edge Deployment

Practical WPT applications require learning models that can operate on resource-constrained embedded platforms. Model compression, quantization, knowledge distillation, and compact network architectures may reduce memory use and computational cost. Lookup-table-based adaptive control [10] and symbolic-regression-based analytical modeling [89] illustrate potential routes toward lightweight implementation. Future studies should report model size, inference time, hardware platform, and control-period requirements and should compare learning-based implementations with conventional controllers and analytical estimators.

### 4) Multi-Agent and Distributed Coordination

WPT systems involving multiple transmitters, receivers, or mobile charging platforms require coordinated decisions across distributed devices. Multi-agent reinforcement learning could support transmitter activation, charging scheduling, beam coordination, and power allocation. Existing single-agent studies on multimodal UAV charging [43] and joint UAV trajectory and resource optimization [74] provide a basis for this development but do not yet demonstrate distributed multi-agent coordination. Future work should address scalability, communication overhead, partial observability, and coordination stability. Federated learning may provide an additional mechanism for collaborative model training without directly exchanging raw operational data.

### 5) Standardized Datasets and Evaluation Protocols

The lack of common datasets and evaluation protocols limits reproducible comparison among learning-enabled WPT methods. Future benchmark datasets should include representative variations in coupler geometry, operating frequency, load, misalignment, temperature, transmission medium, and measurement noise. Evaluation should cover not only prediction or control accuracy but also system efficiency, robustness, inference time, sensor requirements, and deployment cost. Common training-test partitions and cross-device validation protocols would provide a more reliable basis for comparing supervised learning, reinforcement learning, and hybrid methods.

### 6) Digital Twins and Safe Simulation-to-Reality Transfer

Digital twins can provide a controlled environment for data generation, model training, rare-condition evaluation, and controller testing before deployment on high-power hardware. An EV-WPT framework has already integrated genetic-algorithm-based optimization, ANN surrogate modeling, and proximal-policy-optimization-based control within a digital-twin simulation environment [11]. Future research should improve model calibration, online updating, uncertainty representation, and validation against experimental systems. Digital twins should be regarded as an intermediate platform connecting multi-physics simulation, learning-based optimization, hardware-in-the-loop verification, and physical deployment rather than as a replacement for experimental validation.

## 6. Conclusion

This review has systematically examined supervised learning, reinforcement learning, and hybrid or physics-informed methods across three key WPT tasks: structural optimization, parameter and state estimation, and control and optimization. A task-oriented comparison framework was established to relate each learning category to its data requirements, physical-model dependence, computational efficiency, adaptability, interpretability, and engineering readiness. The reviewed studies show that supervised learning is effective for surrogate modeling, parameter regression, and fast inference when representative labeled data are available; reinforcement learning is advantageous for sequential control, adaptive tuning, and scheduling under changing operating conditions; and hybrid or physics-informed methods are promising for combining data-driven flexibility with physical consistency, especially in data-scarce or multi-objective settings. Machine learning has consequently demonstrated clear potential to accelerate electromagnetic design, enable sensorless parameter and position estimation, and improve adaptive WPT control. Practical deployment, however, remains constrained by limited dataset diversity, weak cross-device generalization, safety and interpretability concerns, and the gap between offline simulation and real hardware. Future work should prioritize open benchmark datasets and unified evaluation protocols, deeper integration of physical knowledge and digital twins for safe sim-to-real transfer, lightweight edge implementation, and multi-agent coordination for dynamic multi-transmitter and multi-receiver scenarios.